# Wideband 67-116 GHz cryogenic receiver development for ALMA Band 2

**P. Yagoubov*[1], T. Mroczkowski[1], L. Testi[1], C. De Breuck[1], A. Gonzalez[2], K. Kaneko[2], Y. Uzawa[2], R. Molina[3], V. Tapia[3], N. Reyes[3], P. Mena[3], M. Beltrán[4], R. Nesti[4], F. Cuttaia[5], S. Ricciardi[5], M. Sandri[5], L. Terenzi[5], F. Villa[5], A. Murk[6], M. Kotiranta[6], W. McGenn[7,8], D. Cuadrado-Calle[7,8], G. A. Fuller[8], D. George[7], J.-D. Gallego[9], I. Lapkin[10], M. Fredrixon[10], V. Belitsky[10]**

*[1]European Southern Observatory (ESO), Garching, Germany*
***Email:** pyagoubo@eso.org*
*[2]National Astronomical Observatory of Japan (NAOJ), Mitaka, Tokyo, Japan*
*[3]University of Chile (UCh), Santiago, Chile*
*[4]Istituto Nazionale di Astrofisica (INAF/OAA), Arcetri, Italy*
*[5]Istituto Nazionale di Astrofisica (INAF/OAS), Bologna, Italy*
*[6]Institute of Applied Physics, University of Bern, Switzerland*
*[7]School of Electrical and Electronic Engineering, University of Manchester (UoM), Manchester, UK*
*[8]School of Physics & Astronomy, The University of Manchester (UoM), Manchester, UK*
*[9]Observatorio de Yebes, Guadalajara, Spain*
*[10]GARD, Chalmers University of Technology, Gothenburg, Sweden*

## INTRODUCTION

The Atacama Large Millimeter/sub-millimeter Array (ALMA[1]) is already revolutionising our understanding of the Universe. However, ALMA is not yet equipped with all of its originally planned receiver bands, which will allow it to observe over the full range of frequencies from 35-950 GHz accessible through the Earth's atmosphere. In particular Band 2 (67-90 GHz) has not yet been approved for construction. Recent technological developments in cryogenic monolithic microwave integrated circuit (MMIC) high electron mobility transistor (HEMT) amplifier and orthomode transducer (OMT) design provide an opportunity to extend the originally planned on-sky bandwidth, combining ALMA Bands 2 and 3 into one receiver cartridge covering 67-116 GHz.

The IF band definition for the ALMA project took place two decades ago, when 8 GHz of on-sky bandwidth per polarisation channel was an ambitious goal. The new receiver design we present here allows the opportunity to expand ALMA's wideband capabilities, anticipating future upgrades across the entire observatory. Expanding ALMA's instantaneous bandwidth is a high priority, and provides a number of observational advantages, including lower noise in continuum observations, the ability to probe larger portions of an astronomical spectrum for, e.g., widely spaced molecular transitions, and the ability to scan efficiently in frequency space to perform surveys where the redshift or chemical complexity of the object is not known a priori. Wider IF bandwidth also reduces uncertainties in calibration and continuum subtraction that might otherwise compromise science objectives.

Combined with an upgrade of the ALMA digital processing hardware and software system, the new wideband receiver we describe here will address two of the three key science drivers in the ALMA Development Roadmap [1], a document summarising the highest priority upgrades to enable new ALMA science in the 2030's and beyond. Specifically, this system will help probe the origins of galaxies by measuring highly redshifted lines in broad spectral surveys, and the ability trace the evolution of simple to complex organic molecules in the Milky Way Galaxy, as this portion of the spectrum provides one of the least crowded portions of the spectrum for probing ground state molecular transitions [2].

Here we provide an overview of the component development and overall design for this wideband 67-116 GHz cryogenic receiver cartridge, designed to operate from the Band 2 receiver cartridge slot in the current ALMA front end receiver cryostat.

## RECEIVER DESIGN

The receiver design is based on, and is thus similar to, the currently-fielded ALMA cryogenic cartridges, but has only two cooled stages, with operating temperatures of 15 K and 110 K. The 4 K stage is not used, in contrast to the SIS based ALMA receivers. The 300 K plate hosts the DC bias and WR-10 waveguide vacuum feedthroughs and electrostatic discharge (ESD) protection boards. The 110 K stage is used to heatsink the RF waveguides[2] and DC wiring. The receiver feedhorn, orthomode transducer (OMT), and cryogenic low noise amplifiers (LNA) are all located on the 15 K stage. A rendering of the CAD model for the receiver is shown in Fig.1.

---

[1] ALMA is a partnership of ESO (representing its member states), NSF (USA) and NINS (Japan), together with NRC (Canada), MOST and ASIAA (Taiwan), and KASI (Republic of Korea), in cooperation with the Republic of Chile. The Joint ALMA Observatory is operated by ESO, AUI/NRAO and NAOJ.

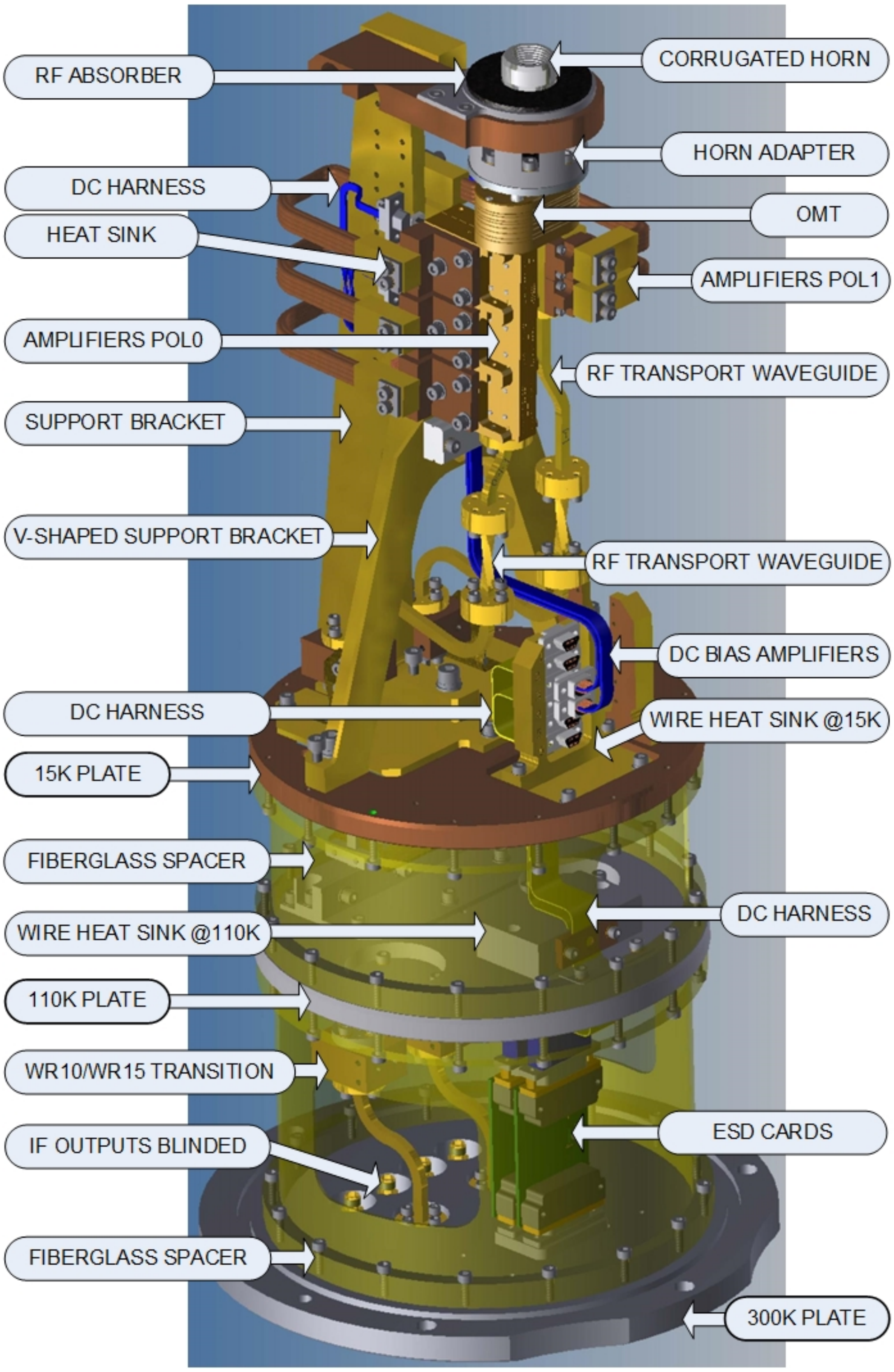

Fig.1. CAD model of a dual-polarisation wideband 67-116 GHz cryogenic receiver.

[2] Here RF stands for radio frequency, and refers to the on-sky frequencies of 67-116 GHz.

**Optics**

The first element in the receiver optics chain is the lens, which also serves as a vacuum window of the cryostat. Since the lens is at room temperature, its insertion loss, beam truncation (at the rim of the lens holder and window, for example), and unwanted reflections contribute significantly to the receiver noise. The baseline material for the lens is currently high-density polyethylene (HDPE) due to its good mechanical properties, relatively low dielectric constant and loss tangent, and the ease of manufacturing, including the anti-reflection (AR) profile. However, such HDPE lens still contributes significantly to the total receiver noise. For that reason we are also investigating ultra-high molecular weight polyethylene (UHMW-PE) and high resistivity float zone silicon as alternative materials which have a potentially lower loss tangent than HDPE (see Chesmore et al. in these proceedings) and allow a more accurate machining of the double layer AR profile. This ongoing design study for the lens optimisation includes accurate loss measurements of the different materials in free space resonators, manufacturing trials with direct and laser machining, and the numerical optimisation of the AR geometry which accounts for the limitations of the manufacturing process.

The next element in the optics is the feedhorn, which is placed on the 15K stage of the receiver. Three prototypes have been developed and tested by University of Chile (UCh) [3], INAF [4] and NAOJ [5]. All the designs are based on corrugated horns with different profiles and have been optimised to achieve low cross-polarisation (<-30 dB), high reflection loss (>30 dB), good beam symmetry and appropriate beam size and phase centre location (PCL) as a function of frequency.

The final optical element is the OMT, which is located after the feed horn to separate the orthogonal linear polarisations fed to different LNAs. The main performance drivers of the OMT are summarised in Table 1. Three versions of the OMT have been developed and manufactured by UCh, INAF, and NAOJ. Fig.2 provides photos of the OMT hardware.

Table 1: OMT requirements

| Specification | Value | Unit | Specification | Value | Unit |
|---|---|---|---|---|---|
| Bandwidth | 67-116 | GHz | Isolation | > 30 | dB |
| Return loss | > 15 (> 20 goal) | dB | Cross-polarisation maximum | < -30 | dB |
| Insertion loss | < 0.5 | dB | Output waveguides | WR10 | NA |

The OMT designed by UCh is based on a turnstile junction architecture [6]. It has a circular waveguide input and two WR-10 rectangular waveguide outputs. The design comprises a turnstile junction, waveguide bends, steps and power combiners. Each of these elements was optimised separately using finite element (FE) simulations. The OMT was constructed in five metal plates fabricated in a CNC milling machine.

The INAF OMT has been designed based on the Turnstile junction and fabricated with platelet technology. The plates were subtractively manufactured using standard machining techniques [7].

The NAOJ OMT design is based on a standard dual-ridged waveguide Boifot junction modified for wideband performance [8]. The design has been performed by the hybrid Mode-Matching/Finite-Elements code in WaspNET, and confirmed by HFSS FE simulations.

The S-parameters of the three OMTs have been measured at INAF, Arcetri, the results are shown in Fig.3. The new prototypes of the UCh and INAF OMTs are being currently developed to minimize the insertion loss through the selection of materials and also improvement of surface roughness.

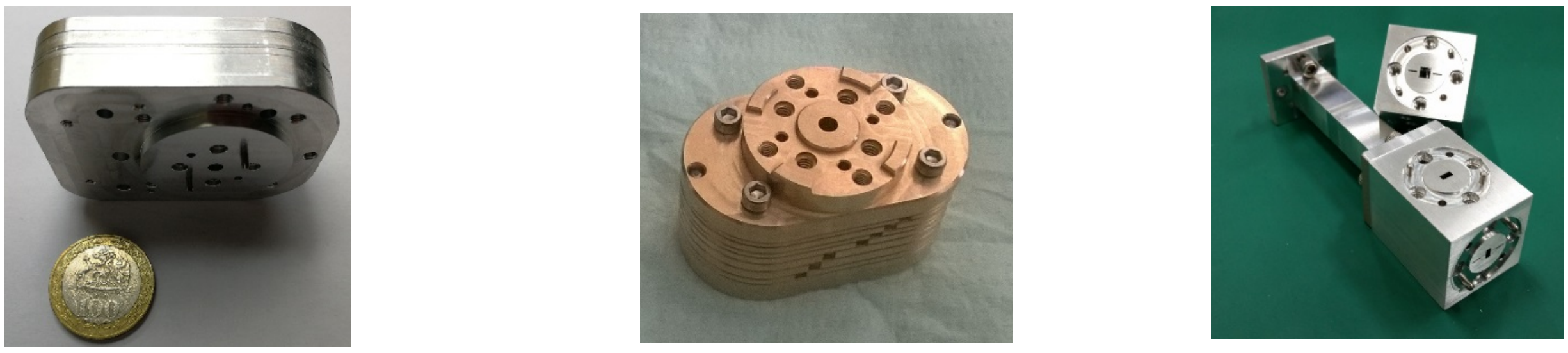

Fig.2. OMT versions of UCh (left panel), INAF (middle panel) and NAOJ (right panel). A 100 Chilean Peso coin (23.5 mm) is shown for scale in the left panel.

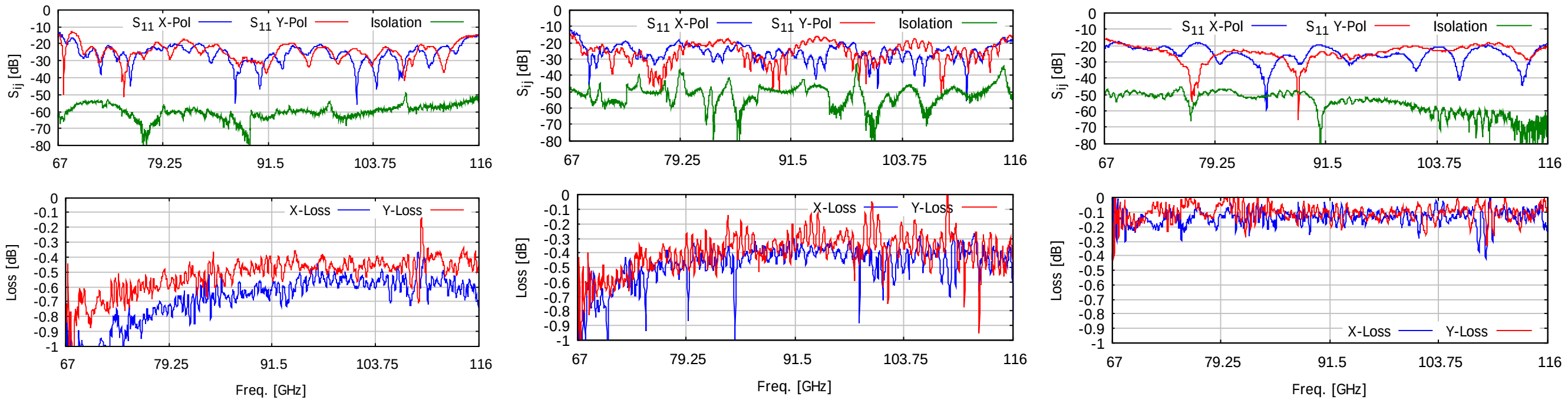

Fig. 3. The measured S-parameters of the UCh (left panel), INAF (middle panel) and NAOJ (right panel) OMTs.

## Low Noise Amplifiers

Two types of low noise amplifier (LNAs) are currently used in the receiver, in different polarisation channels – one developed by the University of Manchester (UoM) and the other commercially produced by the Low Noise Factory (LNF).

The UoM LNA uses the 35nm gate length Indium Phosphide (InP) semiconductor process from Northrop Grumman Corporation (NGC) [9]. The MMIC design consists of two 2-finger transistor stages and features the possibility of independently biasing the gate and drain of each transistor stage. The MMICs are packaged in highly integrated blocks together with waveguide-to-microstrip transitions and bias protection circuit. The packaged LNAs have been characterised at room temperature for S-parameter and noise performance, and at a cryogenic ambient temperature of 20 K for noise performance [10].

The gate length of the LNF InP HEMT process is 0.1 μm [11]. Each MMIC cascades four stages. The 2-finger 40 μm HEMT is used for each stage but connects to different lengths of shorted stubs at source terminal. The source inductive stubs function as negative feedback element that stabilises the device and also brings the optimal impedances for minimal NF and maximal available power gain close to each other, although such inductive feedback decreases the available power gain as well. The MMICs are packaged in compact split block modules with E-plane radial probes to couple the signal from waveguide into the MMIC.

Before installing in a prototype receiver, the noise temperatures of all LNAs were tested at the Yebes Observatory using the same test cryostat at 15 K physical temperature and the same noise measurement system and method. This approach avoids systematic errors from different measurement systems that would otherwise bias the test results. The measurements cover the frequency range 70–116 GHz, as 67–70 GHz is not covered by the test equipment used. The absolute accuracy of the noise temperature determinations is estimated to be 1–2 K. The repeatability and relative accuracy between these measurements is better than 1 K. The measured noise temperature as a function of frequency for each measured LNA configuration is shown in Fig.4.

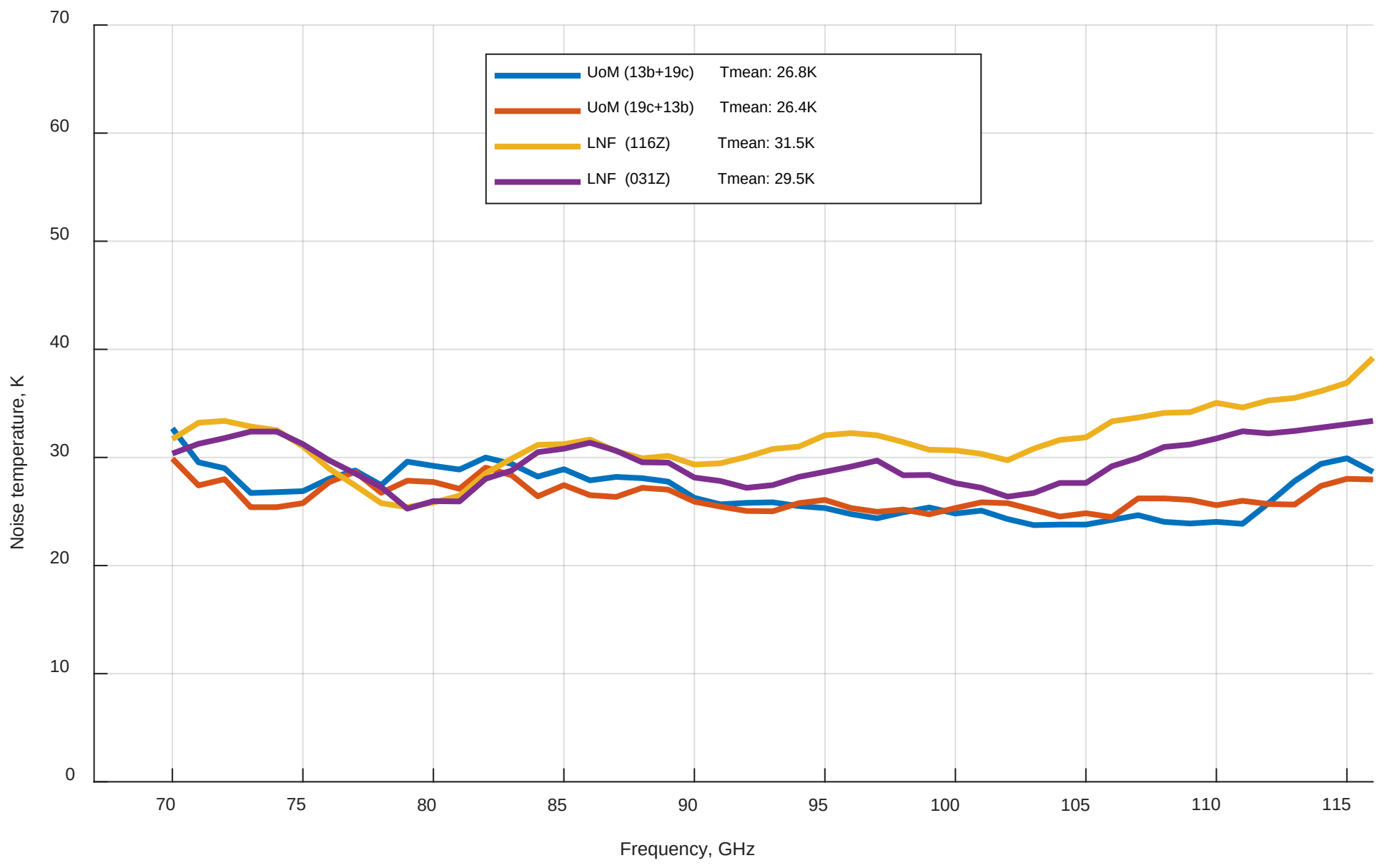

Fig.4. Noise performance of two LNF single block LNAs, and cascaded (2 blocks) UoM LNAs. All LNAs were tested at the Yebes Observatory at 15K ambient temperature, in similar test environment and using the same test procedures.

## TEST SYSTEMS

Optical performance of the system was measured at room temperature in a X-,Y- scanner equipped with Z- rotation stage for co- and cross-polarisation measurements. The maximum scannable plane is 150 mm x 150 mm. The sampling rate in the XY plane is greater than 0.5λ in order to meet the Nyquist criterion. The near-field data are then transformed into far-field data through a Fourier transform.

Other measures of the performance of the receiver, e.g. noise, passband variations, stability etc, were tested using the system shown in Fig.5.

The cryogenic receiver is integrated inside a cartridge test cryostat which is composed of 3 stages, connected with corresponding stages of a GM cryocooler (Sumitomo RDK 3ST), providing references at: ~80K, ~15 K, and ~4K (not used). The downconverter unit is a single channel 2SB receiver developed by RPG [12], slightly modified for noise measurements. For the IF power variations and spurious response measurements, the signals are fed into HP 8564E

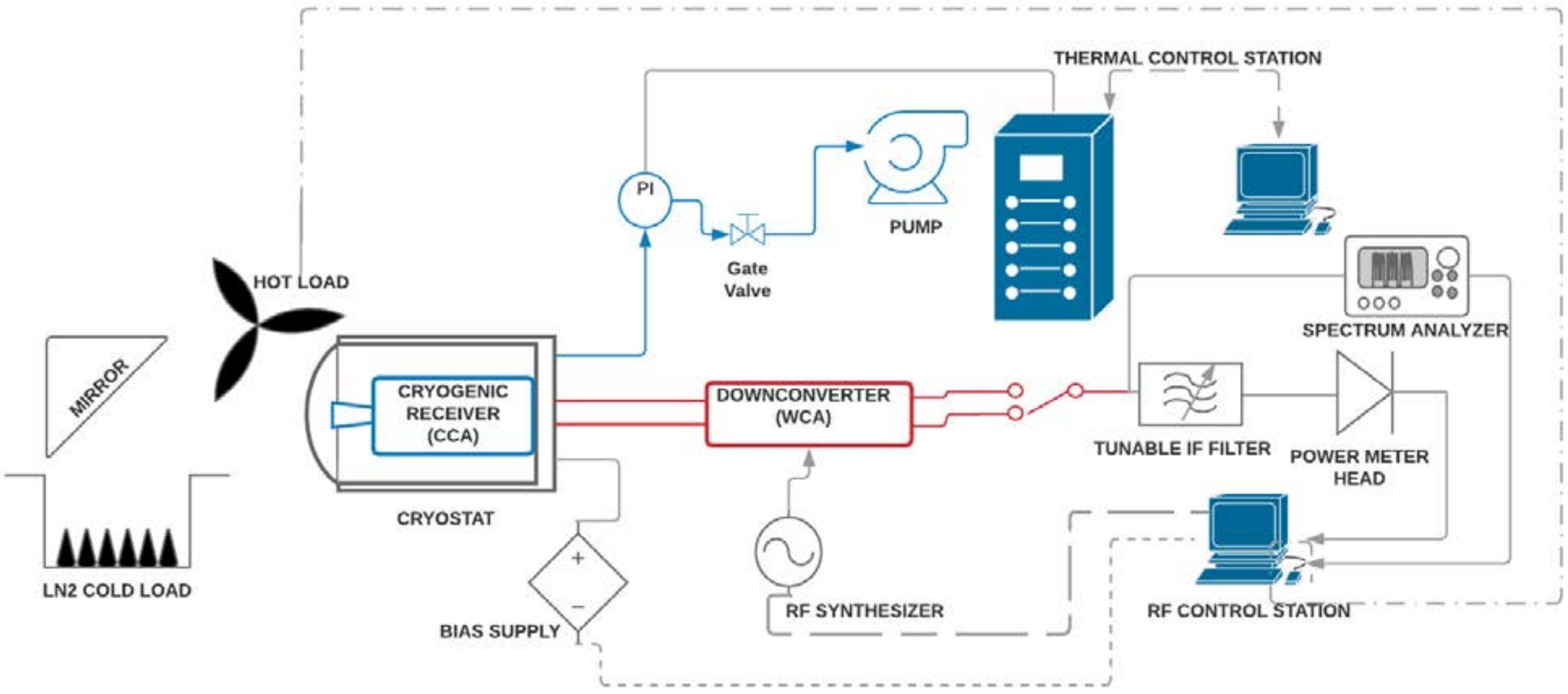

Fig.5. Measurement system block diagram

Spectral Analyzer. An off-the-shelf W-band amplifier from LNF (67 GHz : 116 GHz, ~3dB NF) is added outside the WR-10 rectangular feedthrough flange to increase the total gain by about 25 dB and minimise the noise contribution from the 2SB receiver.

We apply the standard Y-factor method, based on the use of hot and cold thermal loads, to measure the receiver noise. The setup reuses the microwave calibrator produced for the PLANCK-LFI instrument. A panel of Eccosorb CV3 foam absorber (emissivity > 0.999), backed by aluminium and thermalised at room temperature, provides the hot reference: it directly faces the window (lens) of the cryostat, at a distance of a few centimetres, and is connected to a stepper motor allowing the signal to chop between the hot and cold reference. A bed of Eccosorb CR110 pyramids is situated in a large polystyrene box filled by liquid nitrogen (LN2), providing the cold reference. This load (emissivity > 0.9999 at the frequency range of interest) is optically coupled to an aluminium mirror, oriented at 45° w.r.t the beam direction: the size and the position w.r.t the beam direction have been optimised. A fan continuously dries the box and the lens to prevent ice formation on tips and vacuum lens.

## RESULTS

A few prototypes for the optics have been designed, manufactured and tested. The measured beam patterns are in good agreement with electromagnetic (EM) simulations. Fig.6 shows the comparison of the simulated and measured beam patterns of the UCh, INAF and NAOJ designed horn and OMT optical systems, without the lens.

Full optical systems, including the lens, have been measured and optical efficiencies calculated (Fig.7). No data are available for the INAF optics in combination with the silicon lens yet. However earlier measurements with the HDPE lens indicate that the INAF optics performance is very similar. Aperture efficiencies of all optical designs are compliant to ALMA specifications of >80% across full frequency range. Polarisation efficiencies are mostly compliant to the >95.5% requirement as well, with some degradation between 103-107 GHz for all types of optics. We believe this degradation is related to the spurious mode excitation at the feedhorn OMT interface due to possible misalignments. We are currently working on implementing more accurate alignment of these elements and expect improvement in the performance.

The prototype cryogenic receiver was designed to accommodate any version of the optics described above. The first prototype was assembled using the INAF optics. Fig.8 shows the uncorrected receiver noise temperature of the UoM LNAs based polarisation channel. The results agree well with the noise model, based mainly on the measured gain and noise performances of the LNA and downconverter. When no measurement were available for a particular optical component, simulations were used.

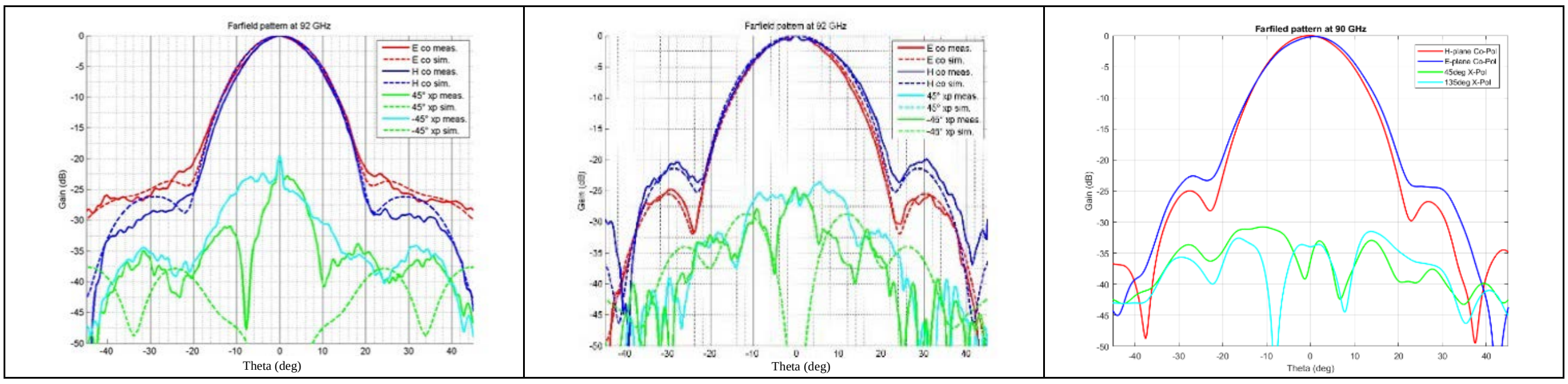

Fig.6. Measured and simulated far-field beam patterns at 92 GHz for UCh horn and OMT (left panel), INAF horn and OMT (middle panel), and at 90 GHz for NAOJ horn and OMT (right panel).

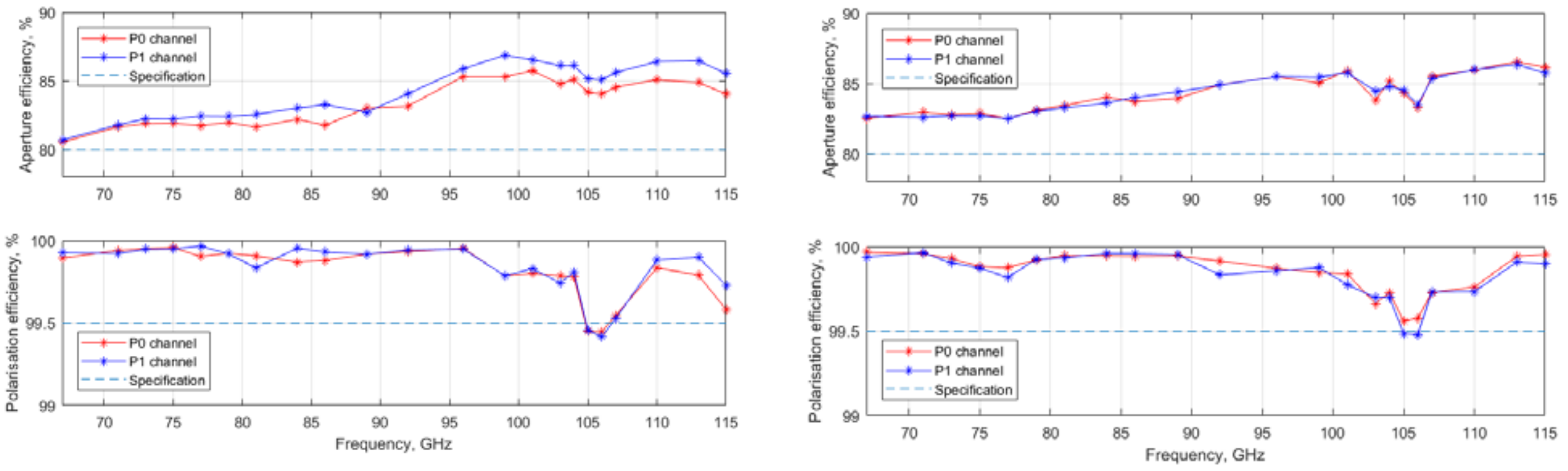

Fig.7. The measured aperture and polarisation efficiencies of the full optics including the Si lens for UCh (left panel) and NAOJ (right panel). Aperture efficiencies are fully compliant with ALMA specification of >80% across full frequency range. Polarisation efficiencies are mostly compliant to ALMA specification of >99.5%, with some degradation between 103-107 GHz.

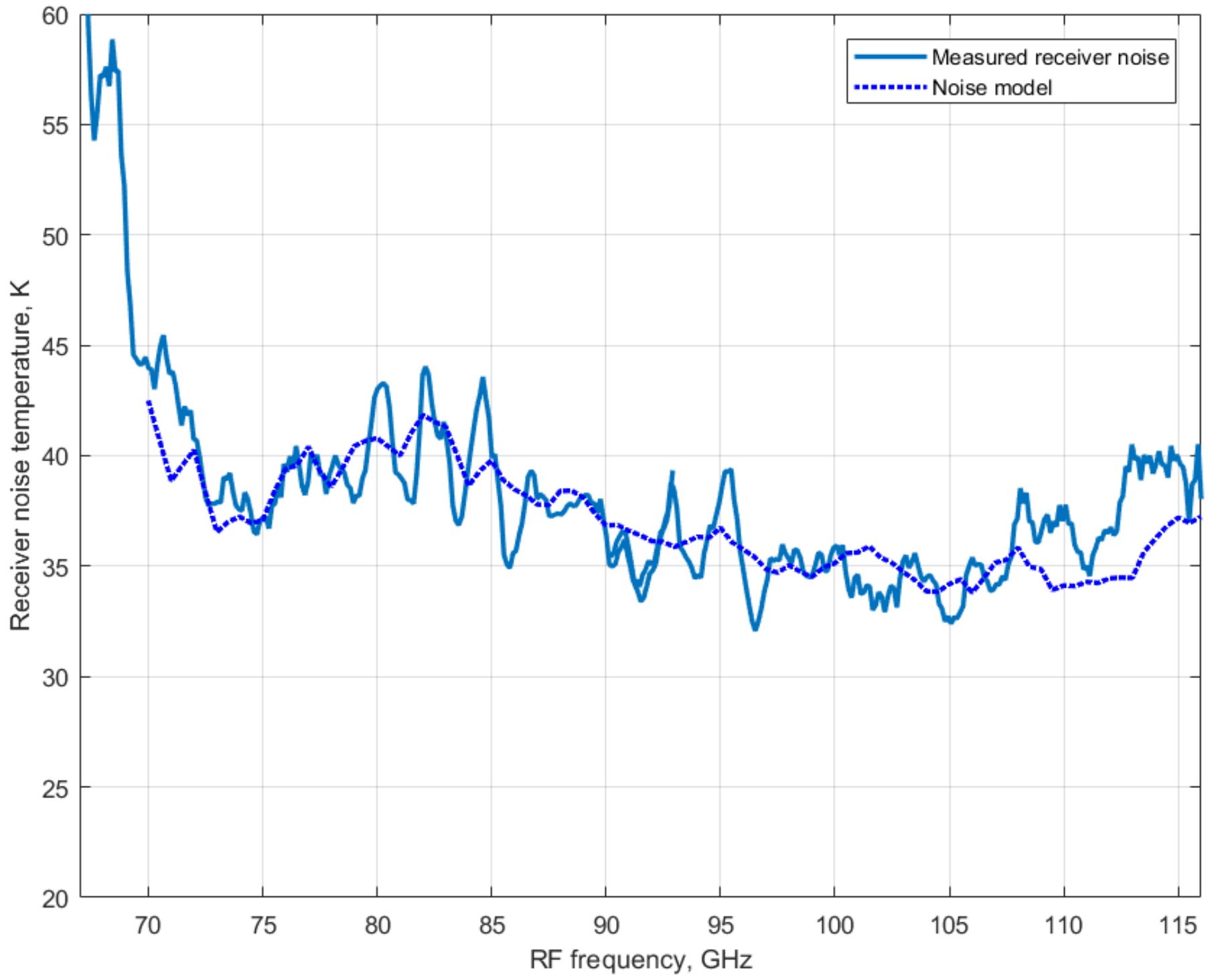

Fig.8. The measured uncorrected receiver noise temperature and comparison to the noise model based on the measured or simulated performances of the LNA and optical elements.

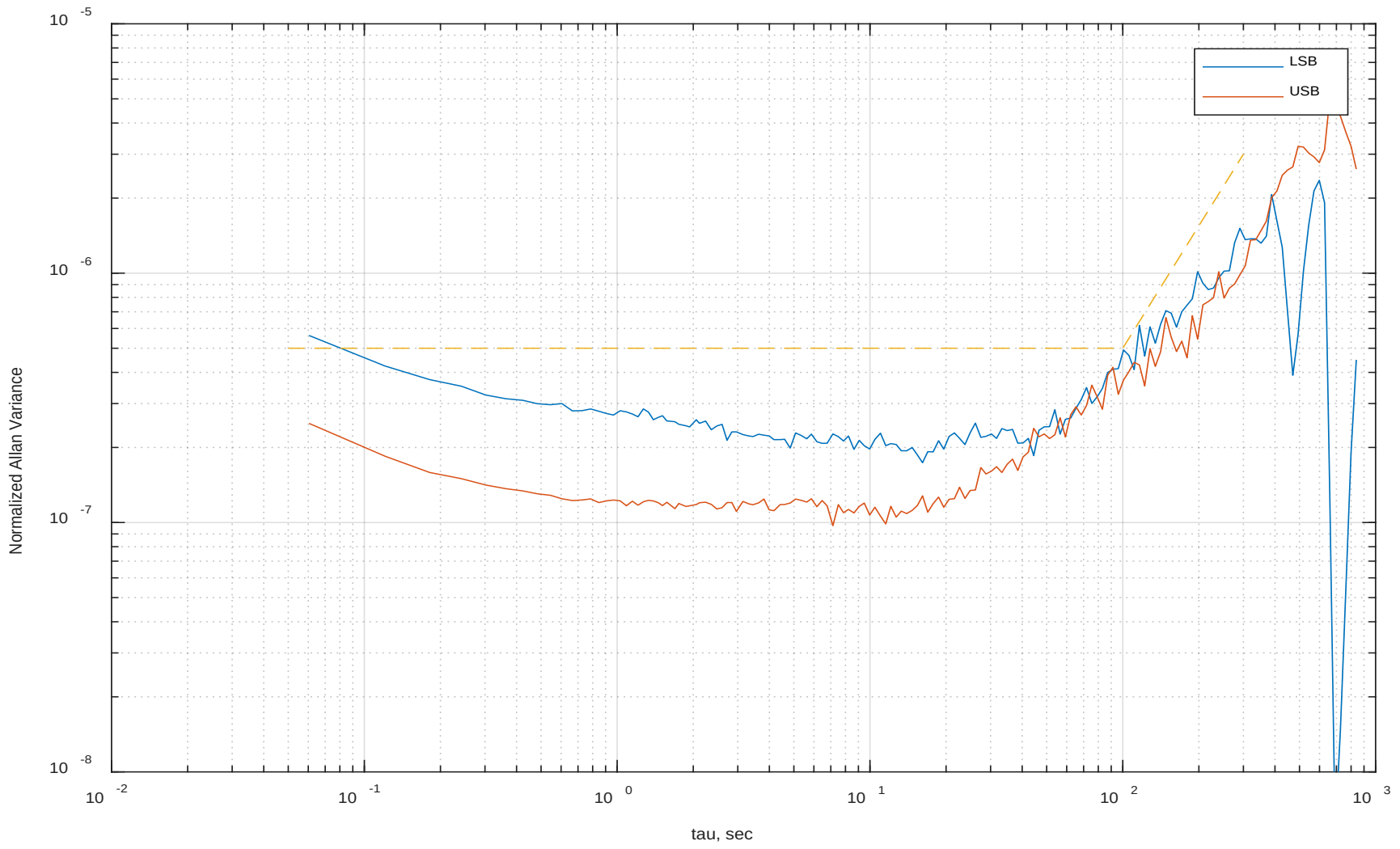

Fig.9. Amplitude stability of the receiver at LO = 100 GHz in two sidebands (USB and LSB). Dashed line indicates ALMA front-end requirement.

Fig.9 shows the measured amplitude stability of the receiver at the LO frequency of 100 GHz in both sidebands. The normalised Allan variance are calculated from the 10 minutes time series measurements.

## CONCLUSIONS

In these proceedings, we have presented an overview of the design for an ALMA Band 2 cryogenic receiver cartridge covering the full 67-116 GHz atmospheric window, originally defined as ALMA Bands 2 and 3. A number of challenges have been met in realising this wideband cartridge design, and we have developed multiple technological solutions to produce critical elements capable of meeting these requirements. This wideband design will open a new window for ALMA, which through ALMA's unparalleled collecting area, resolution, and low noise, will probe astrophysics in this band to much greater sensitivities than currently possible. We expect the new receiver system to improve observational efficiency for a number of key science cases crucial to the ALMA 2030 development vision, laying the foundations for future wide bandwidth receiver systems that more than double ALMA's on-sky bandwidth.